\documentclass[11pt]{article}
\usepackage{lmodern}
\usepackage[utf8]{inputenc}
\usepackage{multirow}
\usepackage{graphicx}
\usepackage{url} 
\usepackage{lineno}
\usepackage{longtable}   
      
\usepackage{placeins}    
\usepackage{booktabs} 
\usepackage{hyperref}
\usepackage[affil-it]{authblk}
\usepackage{soul}
\usepackage{amsfonts}  
\usepackage{booktabs}  
\usepackage[margin=1in]{geometry}
\usepackage{float} 
\usepackage{lscape}
\usepackage[table]{xcolor}
\usepackage{amsmath}

\usepackage{natbib}

\begin{document}
\title{\hrule \vspace{15pt}\textbf{Development of a Data-driven weather forecasting system over India with Pangu-Weather architecture and IMDAA reanalysis Data} \vspace{15pt} \hrule}

\author[1]{Animesh Choudhury}
\author[1,*]{Jagabandhu Panda}

\affil[1]{Department of Earth and Atmospheric Sciences, National Institute of Technology, Rourkela, India}
\affil[*]{Corresponding author: jagabandhu@gmail.com}
\date{}

\maketitle

\begin{abstract}
The current Numerical Weather Prediction (NWP) system has progressed a long way since its inception in the last few decades but still faces many constraints in terms of accuracy, computational efficiency, and scalability. Prediction of weather with a data-driven approach has shown great promise in the recent past, and some of them even outperformed the operational NWP systems. Since these data-driven models are trained on massive amounts of historical weather data, the computational cost of training these is also very high. A regional data-driven weather prediction system can provide a cost-effective way to get weather predictions for a particular region. In this study, a regional data-driven weather forecasting model is developed for the Indian region by efficiently modifying the Pangu-Weather (PW) architecture. The model is trained with the Indian Monsoon Data Assimilation and Analysis (IMDAA) reanalysis dataset with limited computational resources. The model’s ability to predict the weather for the next seven days at 6-hour intervals has been evaluated with Root Mean Square Error (RMSE), Anomaly Correlation Coefficient (ACC), Mean Absolute Percentage Error (MAPE), and Fractional Skill Score (FSS) and found to be encouraging. The prediction results at 6 hours lead time for all variables showed that the MAPE remained below five percent, the FSS values exceeded 0.86, and the ACC was consistently above 0.94, reflecting the model’s overall robustness. Three different prediction approaches (static, autoregressive, and hierarchical) are employed and compared to understand the model's performance with increasing forecasting time. The results demonstrated that the prediction error increases with the increase in lead time for all three approaches. Periodic fluctuation in error metrics, present in the static approach, is absent in the autoregressive approach while visible in the hierarchical approach but with lesser intensity in the predictions after three days. Overall, the hierarchical approach performed the best and had higher computational efficiency. The model’s performance in predicting the cyclone tracks with the hierarchical approach is comparable to both observational and reanalysis datasets. 
\end{abstract}

\begin{center}
    \line(1,0){250}
\end{center}
\clearpage
\section{Introduction}
Since its inception, weather prediction has evolved over the years due to its influence in numerous domains such as agriculture, energy, production, transportation, extreme weather prediction \citep{chen2023swinrdm, budakoti2023transport,kumar2024study}, etc. The researchers have explored different paths to accurately predict the weather. The state-of-the-art weather prediction systems are mainly based on numerical solutions of partial differential equations related to weather variables on a discrete numerical grid \citep{de2023, benjamin2019100}. Efficient computational architecture \citep{alley2019advances}, accurate weather observations \citep{bauer2015quiet,bi2023accurate}, and improved representations of small-scale phenomena \citep{pathak2022fourcastnet} have contributed significantly to the evolution journey of current forecasting systems. Despite their great success, the Numerical Weather Prediction (NWP) systems have certain drawbacks. NWP systems predict complex nonlinear physical processes using weather observations as initial and boundary conditions. The preparation and processing of these input datasets require high computational power. The uncertainties present in the initial and boundary conditions and the higher computational cost limit the accuracy, scalability, and speed of the forecast of these NWP systems \citep{lam2023learning,ben2024,hobeichi2023}.

The data-driven approach has emerged as a potential alternative to NWP. The well-maintained historical weather data \citep{rasp2020weatherbench,choudhury2024bharatbench,nguyen2023climax,shinde2024wxc}, along with the exponential advancement in deep learning techniques and computing architecture, has fueled the use of a data-driven approach to weather prediction. In this approach, neural networks are trained with long-term historical weather data to predict the future atmospheric state using the past atmospheric state as input. In contrast to NWP, this approach does not require any physical understanding of the atmosphere and is significantly faster \citep{pathak2022fourcastnet,nipen2024regional}. The increasing interest in this domain has also encouraged the development of foundational models such as ClimaX \citep{nguyen2023climax}, Aurora \citep{bodnar2024aurora}, and Prithvi-Wxc \citep{schmude2024prithvi}. In recent years, several data-driven models have been developed for global weather prediction \citep{pathak2022fourcastnet,bi2023accurate,lam2023learning,price2025probabilistic,lang2024aifs,chen2023c,chen2023a}, and some of them even outperformed the Integrated Forecasting System (IFS) (Roberts et al., 2018), the operational NWP system developed by the European Center for Medium-Range Weather Forecast (ECMWF). Most of these models are trained on the WeatherBench \citep{rasp2020weatherbench} dataset, which helps to intercompare the models. WeatherBench is prepared especially for the development of data-driven models from the ECMWF ERA5 \citep{hersbach2020era5} reanalysis dataset. The original dataset has a spatial resolution of 0.25 degrees and a temporal resolution of 1 hour. However, some of the models are trained on datasets with lower spatial and temporal resolutions. Since weather prediction is a very complex task, it involves several surface and pressure level variables. Usually, the model architecture contains millions of trainable parameters to perform these complex tasks.  Pangu-Weather (PW) \citep{bi2023accurate}, the first model to outperform the IFS, has around 64 million trainable parameters, and it requires 73000 GPU hours on NVIDIA V100s to train for each lead time. The huge computational resources required to train these models also limit their ablation studies, which again limits the understanding of the role of various model components \citep{to2024architectural}.

PW is built on a 3D Earth-Specific-Transformer (3DEST) architecture and processes surface and pressure-level variables across 13 pressure levels. Another important component of PW is the Earth-Specific-Positional Bias (ESB), which encodes positional information according to Earth’s geometry. The input data is divided into patches and projected into a latent space using a patch embedding technique. The PW architecture adopted a hierarchical encoder-decoder framework derived from the SWIN Transformer \citep{liu2021swin}. This architecture significantly outperforms IFS and FourCastNet \citep{pathak2022fourcastnet}. PW’s extraordinary performance in deterministic and ensemble forecasts makes it an efficient and reliable tool for medium-range weather prediction. In their ablation study, \cite{to2024architectural}To et al. (2024) showed that the two key components of PW, ESB, and 3DEST, are noninfluential to its overall performance and computationally expensive in terms of training time and memory usage. The study also highlights that the Transformer backbone and SWIN mechanism mainly drive the success of PW.

The computational resources required to develop global data-driven weather forecasting systems are mostly available to eminent public research organizations like ECMWF or private entities like Google, Meta, Nvidia, etc. The development of such a sophisticated weather prediction system is not economically viable for low-income nations or research organizations with limited funding. A regional data-driven weather forecasting system can not only reduce the computational burden but also capture regional specificity more prominently. Recently, there has been a growing interest in adopting a data-driven approach for region-specific weather prediction \citep{oskarsson2023graph}. In this study, a medium-range, data-driven weather forecasting system is built using advanced PW architecture and trained on the Indian Monsoon Data Assimilation and Analysis (IMDAA) reanalysis dataset \citep{rani2021imdaa}. The dataset has a higher spatial resolution than ERA5 and is available over India and the surrounding regions. The prediction performance is evaluated with Root Mean Square Error (RMSE), Anomaly Correlation Coefficient (ACC), Mean Absolute Percentage Error (MAPE), and Fractional Skill Score (FSS) and found to be satisfactory. The error statistics provided in this study can act as a baseline for future models trained on the IMDAA dataset over this region. This study represents a pioneering effort to develop a data-driven weather forecasting model for India with limited computational resources. The developed model can provide a cost-effective alternative to traditional NWP for this region. The time and computational cost required by a well-trained data-driven model for predicting weather is negligible compared to that of traditional NWP. The model can be finetuned in the future with more samples, particularly for the prediction of extreme events like cyclones, and can be helpful in developing an early warning system.
 
\section{Method}
\subsection{Reanalysis data}
Systematically collected long-term historical weather datasets are a prerequisite for training a data-driven weather forecasting model. In this study, the IMDAA reanalysis dataset provided by the National Centre for Medium Range Weather Forecasting (NCMRWF) was downloaded from \url{https://rds.ncmrwf.gov.in/}. This dataset provides high-resolution (0.12 degrees) meteorological observation at 24 vertical pressure levels over India and the nearby regions. Four surface variables (10m U, V components of wind, 2m temperature, mean sea level pressure) along with some important pressure level variables (geopotential height, relative humidity, temperature, and the U and V components of wind) were selected for this study. Geopotential height data were considered at four pressure levels (1000, 850, 500, and 50 hPa), while relative humidity, temperature, and the U and V wind components were selected at 850 hPa and 500 hPa. These variables were selected based on the existing literature and available computational power. The study area is bounded by 5°N to 40°N latitude and 65°E to 100°E longitude, which adequately covers the Indian subcontinent and surrounding areas. A detailed summary of the variables and their corresponding pressure levels is provided in Table \ref{tab:Table1}. The short names of the variables mentioned in Table 1 will be used hereafter. The dataset is collected from 1990 to 2020 four times daily (00 UTC, 06 UTC, 12 UTC, and 18 UTC). The whole study period was divided into training (1990-2017), validation (2018), and testing (2019-2020). Due to the limitation of computational resources, 5000 random samples were considered from the training period, while 500 samples were considered for validation. The dataset was normalized with mean and standard deviation before infusing into the model.
\subsection{Computation Framework}
All the models were developed and trained on a system with an 11th Gen Intel Core i7-11700 CPU, featuring 16 logical processors (8 cores with two threads per core) operating at a base frequency of 2.50 GHz, with a maximum turbo frequency of 4.90 GHz. The CPU includes 48 KB of L1d cache, 32 KB of L1i cache, 512 KB of L2 cache, and 16 MB of L3 cache. The model architecture, training pipeline, and result analysis were done with Python 3.11 and Python-based libraries such as PyTorch 2.5.1, Matplotlib, xarray, numpy, pandas, etc. 
\subsection{Model Architecture} 
\begin{figure}[p] 
\includegraphics[width=\textwidth]{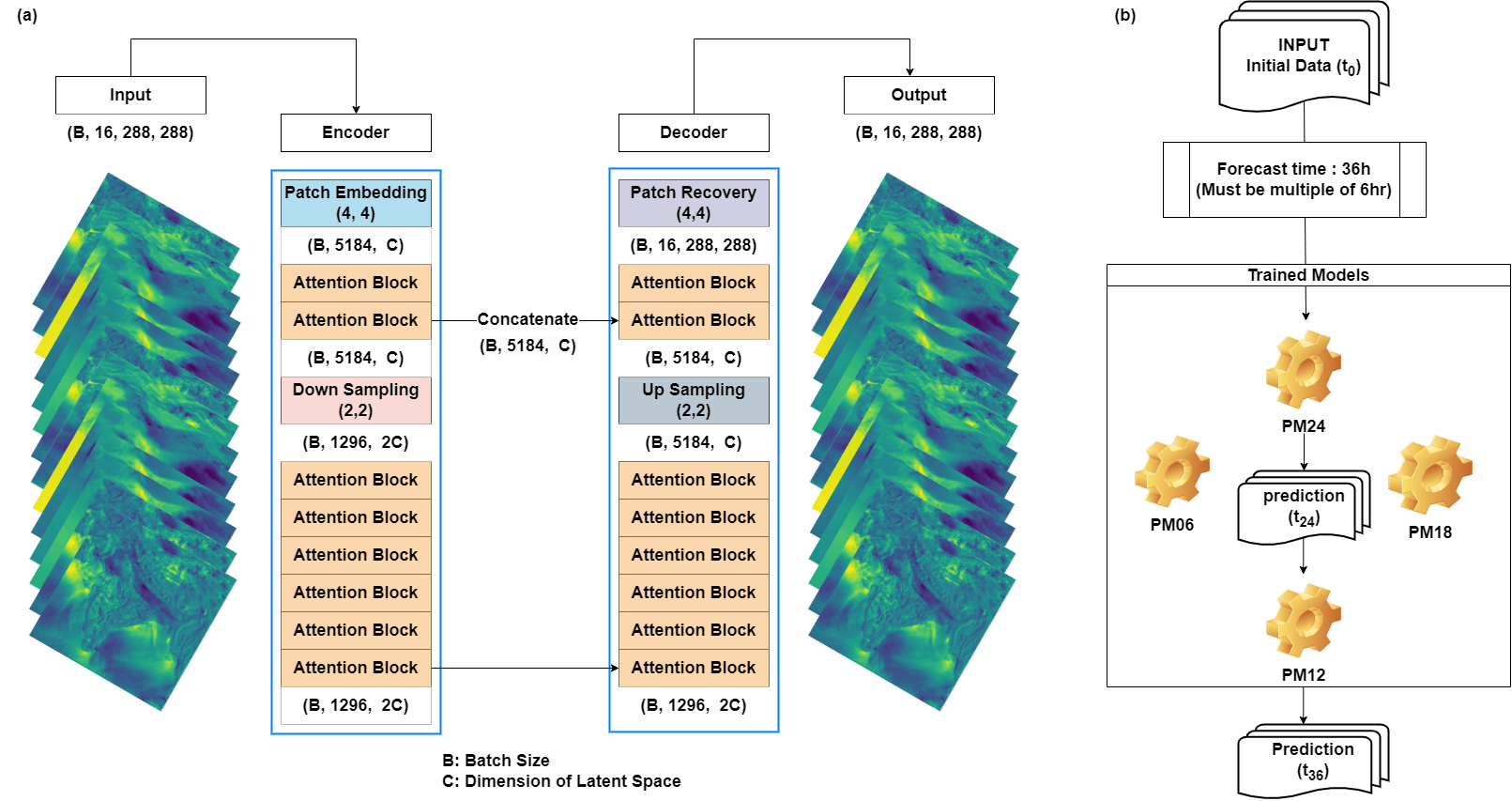}
\caption{(a) Model architecture adopted in this study, showing the data dimension in each block of the model component, (b) example of hierarchical forecasting approach for a prediction of 36 hours. }
\label{fig:Figure1}
\end{figure}

This study adopted the model architecture provided by \cite{bi2023accurate} and further modified it according to the insight provided by \cite{to2024architectural}. The two main components of PW, ESB, and 3DEST, were replaced with simple positional embedding and 2D attention mechanism, respectively, which significantly reduced the requirement of computational power. The model architecture was adjusted as per the dimensions of the input data. The input data had the shape of 16 × 288 × 288, where 16 is the number of channels, which include the surface variables and pressure variables at each pressure variable (Table \ref{tab:Table1}). The data was initially embedded to a latent space with dimension ‘C’ from the actual space. Patch Embedding, a commonly used dimensionality reduction approach, was employed with a patch size of 4 × 4. The stride of the sliding window was the same as the patch size. This embedded data went into a standard encoder-decoder architecture, having eight encoder layers and the same number of decoder layers. The data dimension remained unchanged for the first two layers of the encoder, while in the next six layers, the horizontal dimension was halved, and the channel dimension was doubled. The decoder was the mirror image to that of the encoder. The output of the second layer of the encoder and the seventh layer of the decoder were concatenated along the channel dimension. The study applied SWIN transformers and linked the adjacent layers of different shapes with down-sampling and up-sampling operations. The output from the decoder was then transformed into the original space from the latent space with the help of patch recovery. Both patch embedding and patch recovery had the same number of parameters but were not shared with each other. The flow of the data through the different components of the model is shown in Figure \ref{fig:Figure1}(a).
\subsection{Model training process}
Three different approaches were chosen to evaluate and improve the model’s performance. These were static, autoregressive, and hierarchical. In a static setting, the prediction provided by a model trained for predicting weather 6 hours ahead is compared with the actual observation of a longer duration. This provided a primary baseline for all the variables for different lead times. The model’s performance could be considered improved if it beat this baseline. In the autoregressive approach, the prediction provided by the model is used as input for the next iterative prediction. One of the main drawbacks of this approach is the propagation of error through the prediction loop. The initial prediction error amplifies nonlinearly for a longer prediction time, as observed in conventional NWP systems. Also, the time and computational resources required to make predictions for longer prediction times increase proportionally. For example, a model trained for predicting 6 hours ahead required four iterations to provide a 24-hour prediction. For the hierarchical approach, models for multiple lead times (06, 12, 18, and 24 hours) were trained. This approach prioritizes the use of the deep network with the longest feasible lead time at each step, which consequently reduces the number of iterative forecast steps and propagation errors accumulated due to repeated short-term prediction. For instance, when generating a 36-hour forecast, the hierarchical approach first utilizes the 24-hour forecast model once, followed by a 12-hour forecast once, rather than iteratively applying the 6-hour forecast model 6 times (Figure \ref{fig:Figure1}(b)). To ensure computational efficiency and stability during training, a batch size of 2 was chosen. The training process employed the Adam optimizer with an initial learning rate of 0.0001, and Mean Squared Error (MSELoss) was used as the loss function to minimize prediction errors. To enhance learning adaptability and prevent overfitting, the ReduceLROnPlateau scheduler was employed, with a patience of 3 epochs and a learning rate reduction factor of 0.1. This training configuration allowed the model to converge effectively while balancing accuracy and computational resources.
\subsection{Ablation study on the Latent space dimension}
The memory required for training significantly exceeds the model's size because intermediated states from both forward and backward passes must be stored \citep{rajbhandari2020zero}. By reducing the model size, a larger local batch size can fit into a single GPU, thereby lowering the computational cost and improving training efficiency. While \cite{to2024architectural} conducted a comprehensive ablation study on PW, their analysis maintained a constant C. However, C significantly influences the model's trainable parameters and memory requirements (Table \ref{tab:Table2}). To address this gap, the present study investigates the impact of varying latent space dimensions on model convergence, focusing on training dynamics such as epochs and learning rate. The model was trained with an initial learning rate of 1×10\textsuperscript{-4}, and the training process continued for up to 200 epochs or until the learning rate decayed to 1×10\textsuperscript{-8}, whichever occurred first. Experiments were conducted for C = 12, 24, 48, 96, and 192 to evaluate their effects on convergence behavior and computational efficiency.  
\subsection{Evaluation Matrix}
The performance of a global data-driven weather forecasting system is generally evaluated through latitude-weighted error metrics to account for the positional bias. In this study, the use of traditional error metrics was assumed to be more appropriate as the study area is situated within a small latitude range of the northern hemisphere and to keep the evaluation process simple. To effectively capture the model’s predictive capabilities, Root Mean Squared Error (RMSE), Anomaly Correlation Coefficient (ACC), Mean Absolute Percentage Error (MAPE), and Fractional Skill Score (FSS) were computed between the actual and predicted observations. Each matric provides a unique inside into the model’s ability to predict atmospheric variables across lead times, offering a comprehensive assessment of its accuracy, spatial reliability, and robustness. These evaluation metrics were selected to understand the model’s accuracy, precision, and bias. 

RMSE measures the average magnitude of error between the predicted and actual values. Lower RMSE values indicate better model performance, and it is calculated as:
\begin{equation}
\operatorname{RMSE}(i, j)=\sqrt{\frac{1}{N} \sum_{n=1}^N\left(Y_n(i, j)-\hat{Y}_n(i, j)\right)^2}
\end{equation}
Where $Y_n(i, j)$ and $ \hat{Y}_n(i, j)$ are the actual and predicted values at grid cell $\left(i,j\right)$ for sample n respectively and N is the total number of samples.

ACC measures the correlation between anomalies of predicted and actual values and evaluates the model’s ability to capture spatial and temporal patterns. ACC values range from -1 to 1. A value close to 1 indicates that the model captures the variability and spatial patterns well, while negative values indicate poor performance. In this study, ACC is computed as:
\begin{equation}
\operatorname{ACC}(i, j)=\frac{\sum_{n=1}^N\left(Y_n(i, j)-\bar{Y}(i, j)\right)\left(\hat{Y}_n(i, j)-\overline{\hat{Y}}(i, j)\right)}{\sqrt{\sum_{n=1}^N\left(Y_n(i, j)-\bar{Y}(i, j)\right)^2 \cdot \sum_{n=1}^N\left(\hat{Y}_n(i, j)-\overline{\hat{Y}}(i, j)\right)^2}}
\end{equation}
Where $\bar{Y}(i, j)=\frac{1}{N} \sum_{n=1}^N Y_n(i, j)$ and $\overline{\hat{Y}}(i, j)=\frac{1}{N} \sum_{n=1}^N \hat{Y}_n(i, j)$ are the mean of actual and predicted values at grid cell $\left(i,j\right)$ respectively.

MAPE quantifies the average percentage error between predictions and actual values, making it a relative measure. This is less sensitive to the large error values compared to the RMSE. Lower MAPE values are better, with values below ten percent often considered excellent. MAPE is calculated as:
\begin{equation}
\operatorname{MAPE}(i, j)=\frac{100}{N} \sum_{n=1}^N\left|\frac{Y_n(i, j)-\hat{Y}_n(i, j)}{Y_n(i, j)}\right|
\end{equation}

FSS evaluates the spatial agreement between binary fields of predicted and observed values above a threshold (climatological mean in this case). FSS ranges from 0 (no skill) to 1(perfect skill). Higher values indicate better spatial agreement. Values below 0.5 often indicate poor predictive skill. FSS for each grid cell is calculated as:
\begin{equation}
\operatorname{FSS}(i, j)=\frac{2 \cdot \sum_{n=1}^N\left(P_n(i, j) \cdot O_n(i, j)\right)}{\sum_{n=1}^N P_n(i, j)+\sum_{n=1}^N O_n(i, j)}
\end{equation}
Where $P_n(i, j)=1\left(\hat{Y}_n(i, j)>T\right)$ and $O_n(i, j)=1\left(Y_n(i, j)>T\right)$ are predicted and actual binary field at grid cell $\left(i,j\right)$ for sample n, threshold T. 
\subsection{Tropical cyclone tracking}
To evaluate the model’s performance, the study assessed its ability to predict cyclone tracks by comparing its output with the IMDAA reanalysis dataset and the International Best Track Archive for Climate Stewardship (IBTrACS) \citep{knapp2010international}. The latitude and longitude were collected from the IBTrACS dataset for times 00, 06, 12, and 18 UTC. The reanalysis data 6 hours prior to the initial observation was used as input for the hierarchical prediction. The cyclone’s presence was identified by locating the point of maximum vorticity with a threshold value greater than 5 × 10\textsuperscript{-5} and verifying the presence of a local minimum of mslp within a five-degree radius. The cyclone’s position was determined by tracking the local minimum mslp in the predicted data. The tracking error was estimated using the Haversine formula \citep{winarno2017location}.

\section{Results and discussions}
\subsection{Selection of optimal value of ‘C’}
\begin{figure}[p]
\includegraphics[width=\textwidth]{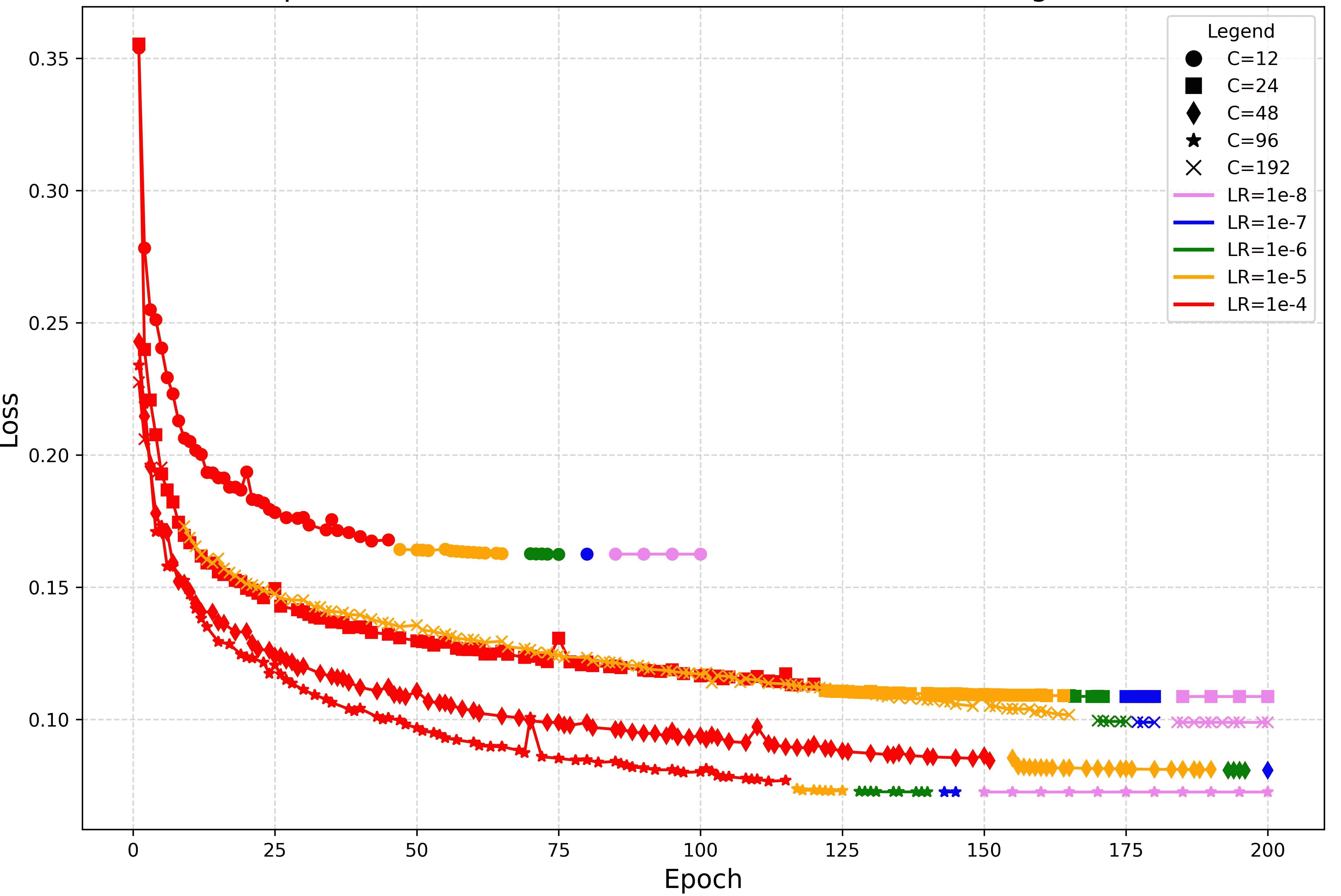}
\caption{Learning curve of the model for different values of latent space dimension (C) shown with different symbols. The different colours highlight the different learning rates..
}
\label{fig:Figure2}
\end{figure}

For all tested values of C, the loss function exhibited a rapid decline during the initial epochs, indicating effective learning and convergence (Figure \ref{fig:Figure2}). This behavior is typical in deep learning models, where the initial phase of the training captures the most significant patterns in the data. However, as training progressed, the rate of improvement in model performance began to slow. To address this, the learning rate was reduced if there was no improvement for three consecutive epochs. The results revealed a general trend where higher latent space dimensions achieve lower loss values. However, this trend was not strictly monotonic. While the largest dimension tested (C = 192) was expected to perform best due to its greater number of trainable parameters, its loss curve closely resembled that of the much smaller dimension (C = 24). In contrast, C = 48 and C = 96 demonstrated superior performance, with C = 96 achieving the lowest overall loss (Figure \ref{fig:Figure2}). This optimal balance between model complexity and generalization capability led to the selection of C = 96 for further experimentation. The observed behavior highlights the importance of carefully selecting latent space dimensions. While higher dimensions can theoretically capture more intricate features, they also increase computational costs and the risk of overfitting. The results underscore the need for empirical evaluation to identify the optimal configuration that balances model capacity, training efficiency, and predictive performance.
\subsection{Evaluation of the model’s performance}
In this study, predictions were compared with the actual observation for the next seven days (168 hours) at 6-hour intervals to understand the model’s near-future prediction capability. The model’s performance was evaluated with RMSE, ACC, MAPE, and FSS for the testing years. The prediction for six hours ahead is the same for all the adopted prediction approaches. This provides baseline errors and highlights inherent limitations present within the model’s architecture. The model demonstrated strong predictive capability for all selected atmospheric variables 6 hours ahead (Table \ref{tab:Table3}). The results highlight the challenge of accurately forecasting wind components at the surface (10m) and higher pressure levels (850 hPa and 500 hPa). The performance in predicting the surface temperature was found to be worse than that of the higher-level temperature. Across all variables, the MAPE remained below five percent, the FSS values exceeded 0.86, and the ACC was consistently above 0.94, reflecting the model’s overall robustness. These results indicate that while certain variables, such as wind speed and relative humidity, may require further refinements to improve local-scale accuracy, the model performed exceptionally well in forecasting large-scale meteorological features 6 hours ahead, making it a reliable tool for short-term weather prediction. The result provided in Table \ref{tab:Table3} can be used as a benchmark for future model development. 

\begin{figure}[p]
\includegraphics[width=\textwidth]{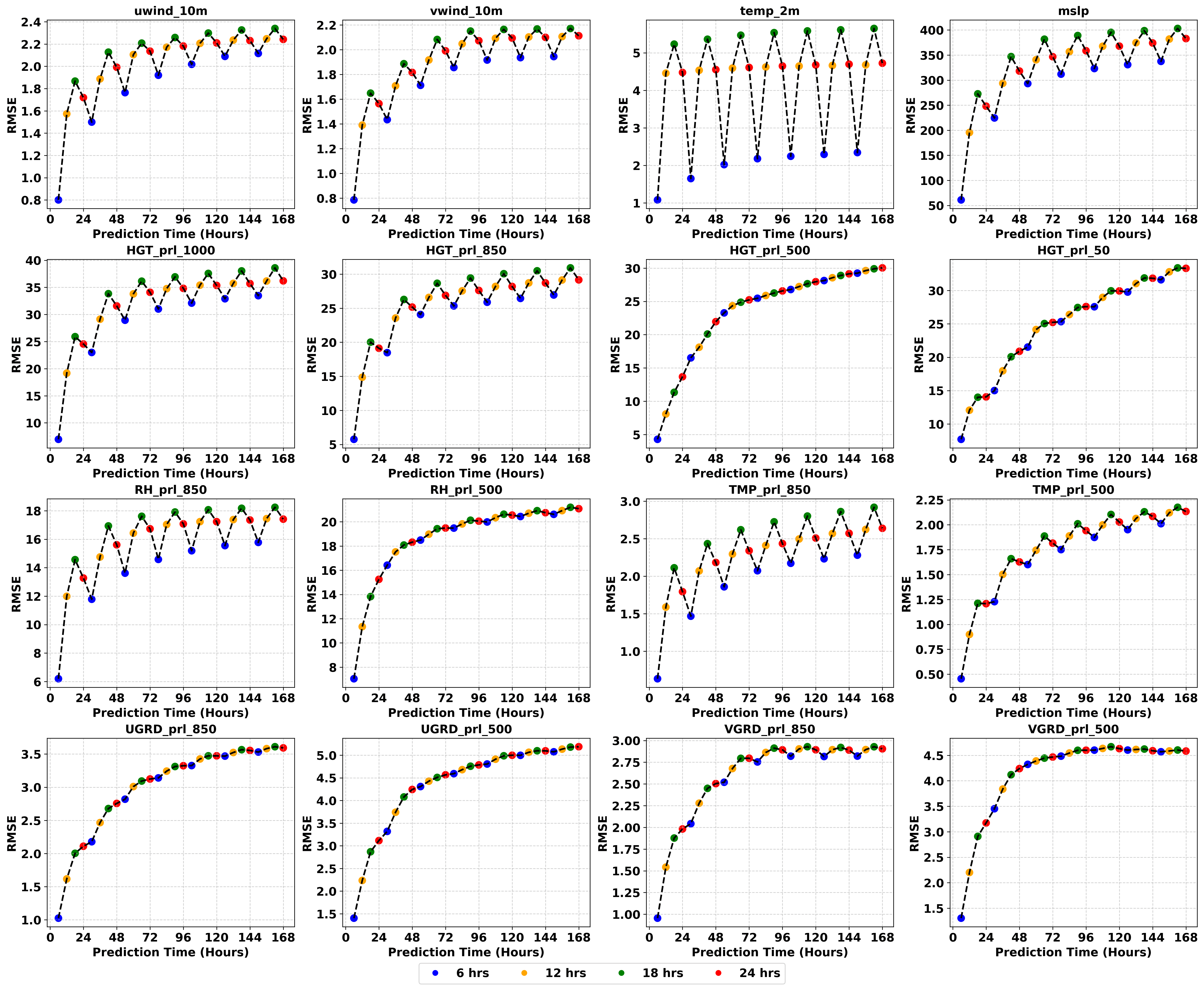}
\caption{RMSE in predicting each variable for the next seven days with a static forecasting approach. The colors blue (06 hours), orange (12 hours), green (18 hours), and red (24 hours) represent the initial prediction and subsequent forecast at 24-hour intervals.
}
\label{fig:Figure3}
\end{figure}

\begin{figure}[p]
\includegraphics[width=\textwidth]{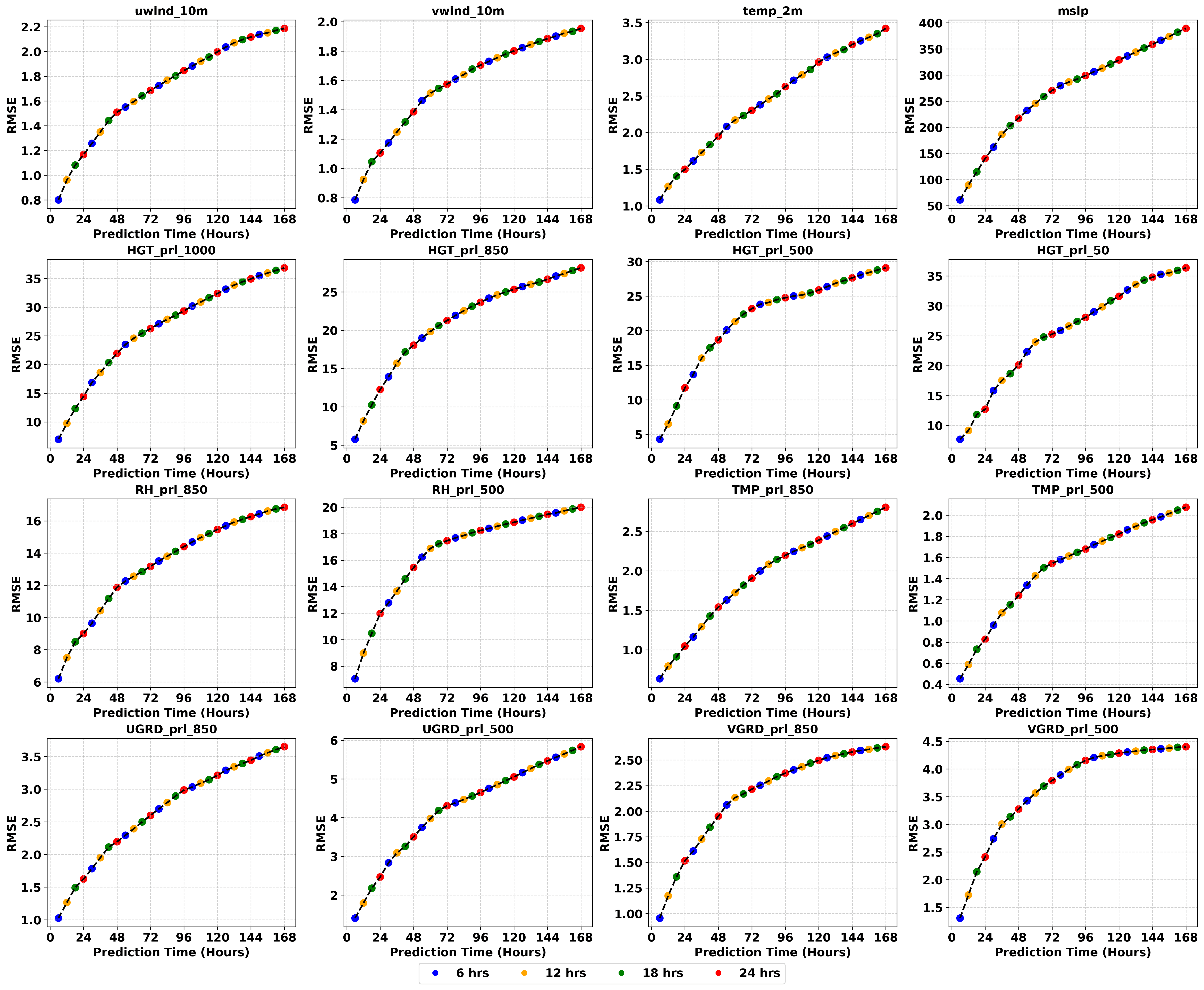}
\caption{RMSE in predicting each variable for the next seven days with an autoregressive forecasting approach. The colors blue (06 hours), orange (12 hours), green (18 hours), and red (24 hours) represent the initial prediction and subsequent forecast at 24-hour intervals..
}
\label{fig:Figure4}
\end{figure}

\begin{figure}[p]
\includegraphics[width=\textwidth]{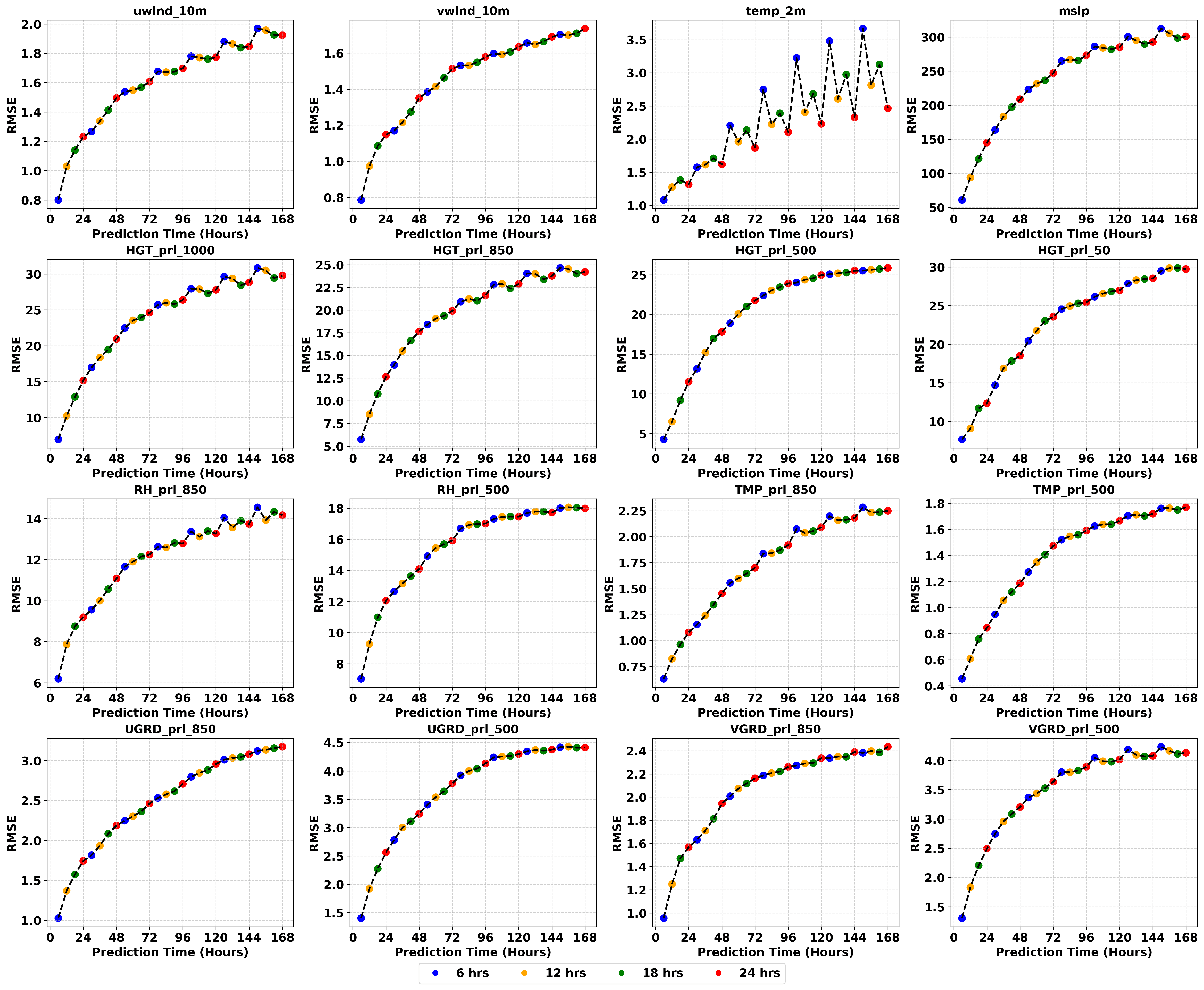}
\caption{RMSE in predicting each variable for the next seven days with a hierarchical forecasting approach. The colors blue (06 hours), orange (12 hours), green (18 hours), and red (24 hours) represent the initial prediction and subsequent forecast at 24-hour intervals.
}
\label{fig:Figure5}
\end{figure}

The static approach showed periodic fluctuations in the error matrices in most of the variables due to the influence of diurnal cycles on forecasting dynamics (Figure \ref{fig:Figure3}). RMSE increased non-linearly with lead time, consistent with the chaotic nature of atmospheric dynamics, where minor initial errors amplified over time. The error growth over time for most variables remained gradual and controlled. Higher RMSE and lower ACC in surface variables highlighted the difficulty in predicting small-scale phenomena that are influenced by local factors like topography, local winds, and diurnal variations. The higher accuracy in predicting upper-level variables suggested that the model benefited from the smoother gradients and larger scales. MAPE and FSS further approved the model’s performance. Next, the study explored the autoregressive approach and observed a reduction in RMSE (Figure \ref{fig:Figure4}) and an improvement in ACC. The periodic fluctuations observed during the static approach were completely absent. The increase in RMSE and decrease in ACC with the increase in lead time was relatively smooth and gradual. Finally, the study investigated the performance of the hierarchical approach in the model’s prediction capability. The results showed a better overall performance compared to the static and autoregressive approach. The lower RMSE (Figure \ref{fig:Figure5}) values and higher ACC values confirmed its superiority over the other selected approaches. In some of the variables, the periodic fluctuation started appearing in the predictions after three days but not with the same intensity as the static approach. The comparison of RMSE of three different approaches in predicting three and five days ahead is presented in Table \ref{tab:Table4}.  All the matrices were consistent and complementary for different prediction approaches. ACC, MAPE, and FSS for different approaches were provided in the supplementary figures (S\ref{fig:S1},\ref{fig:S2},\ref{fig:S3},\ref{fig:S4},\ref{fig:S5},\ref{fig:S6},\ref{fig:S7},\ref{fig:S8},\ref{fig:S9}). This comprehensive evaluation not only validates the model’s ability for forecasting but also identifies areas for improvement.
 
\subsection{Cyclone track prediction}
For cyclone tracking, the study employed a hierarchical prediction strategy to generate weather forecasts at 6-hour intervals. Four cyclones, namely Fani, Bulbul, Amphan, and Nivar, which occurred between 2019 and 2020, were selected for analysis based on their intensity and track characteristics. The results demonstrated that the model accurately predicted the cyclone tracks, with performance comparable to both observational and reanalysis datasets (Figure 6). The average error between the model’s predictions and observed tracks was 132 km, while the error can be partially attributed to inherent errors in the reanalysis datasets, as the average difference between observed and reanalysis tracks was around 77 km. These findings highlight the model’s capability to reliably track cyclones while also underscoring the influence of input data quality on prediction accuracy.
\section{Limitations and future scope}
While the study provides valuable insights into the development and performance of a data-driven weather prediction system, several limitations must be acknowledged. Future studies should focus on addressing these limitations and provide an improved understanding of the same. The study is focused on a specific geographic region. The value C for this location may differ for other locations and need further investigation. Regional models may not adequately capture influences coming from the outside, such as global or remote atmospheric processes that could impact local weather patterns. The model performed well comparably for upper-level atmospheric variables, which proves its strength in capturing large-scale, deterministic atmospheric features. However, its performance for surface-level variables is less consistent, as evidenced by higher RMSE and MAPE values, as well as lower ACC and FSS scores. 
The current model architecture operates at a fixed spatial-temporal resolution and incorporates a limited number of variables and pressure levels. While this simplification facilitates computational efficiency, it may compromise the model’s ability to capture finer-scale atmospheric processes. To address the challenges, future models could incorporate longer-term higher-resolution data, advanced physical parameterizations, and machine-learning techniques tailored to small-scale phenomena. Additionally, integrating localized observational data or leveraging hybrid approaches that combine data-driven methods with physical modeling could further improve performance. A dedicated model is highly recommended for predicting extreme events over a general weather prediction system. This is because general models are typically trained on a dataset dominated by “normal” weather conditions, which can bias the model toward predicting average or non-extreme outcomes. A specialized model, explicitly trained on extreme event data and incorporating tailored physical constraints or higher-resolution inputs, would be better equipped to identify and predict these high-impact phenomena.
Since no prior models have been developed specifically for this region using similar data, there is no direct basis for comparison with other models. Additionally, NCMRWF does not provide initial conditions, making it impossible to compare the model’s performance with the operational forecasting system. This lack of comparative benchmarks limits the ability to contextualize the model’s performance within the border landscape of weather prediction systems. As this study represents a pioneering effort in the region, future research should focus on establishing standardized benchmarks for comparison. Developing open datasets with initial conditions and encouraging the creation of alternative models will facilitate a more robust evaluation of predictive performance and foster innovation in this field. Future efforts should aim to integrate the model into operational forecasting systems and evaluate its performance in real-time scenarios. This would provide practical insights into its applicability and reliability for operational weather prediction, paving the way for its adoption by meteorological agencies.
\newline
\newline
\textbf{Acknowledgments:} Authors gratefully acknowledge National Centre for Medium Range Weather Forecasting (NCMRWF), Ministry of Earth Sciences, Government of India, for IMDAA reanalysis. IMDAA reanalysis was produced under the collaboration between the UK Met Office, NCMRWF, and IMD with financial support from the Ministry of Earth Sciences under the National Monsoon Mission program.
\newline
\newline
\textbf{Data availability:} This study used a subset of IMDAA dataset downloaded from \url{https://rds.ncmrwf.gov.in/datasets}. The total volume of the dataset is around 240 GB. The ground truth observation of cyclone tracks was downloaded from the IBTrACS project (\url{https://www.ncei.noaa.gov/products/international-best-track-archive}). All the datasets used in this study are available in the public domain for research works. The secondary data can be made available upon request through the proper channel. 
\newline
\newline
\textbf{Code availability:} The base code for the model development is taken from \url{https://github.com/DeifiliaTo/PanguWeather} and modified according to the dataset and research objectives. The study used several Python libraries, which include matplotlib, pandas, numpy, xarray, etc. The trained models and the actual codes can be shared if requested through the proper channel.

\newpage
\small
\bibliography{references.bib}
\bibliographystyle{apalike}

\clearpage  
\renewcommand{\floatpagefraction}{0.0}
\newpage

\begin{landscape}
\begin{table}[p]
\centering
\small
\caption{Selected variables used in this study}
\label{tab:Table1}
\vspace{10pt}
\resizebox{\columnwidth}{!}{%
\begin{tabular}{|l|l|l|l|l|l|}
\hline
 &
  Long Name &
  Short Name &
  Description &
  \begin{tabular}[c]{@{}l@{}}Pressure Levels \{plevel\}\\    \\ (hPa)\end{tabular} &
  Unit \\ \hline
\multirow{4}{*}{Surface Variables} &
  U-Component of Wind &
  uwind\_10m &
  Wind in x/longitude-direction   at 10 m height &
   &
  (ms-1) \\ \cline{2-6} 
 &
  V-Component of Wind &
  vwind\_10m &
  Wind in y/latitude-direction   at 10 m height &
   &
  (ms-1) \\ \cline{2-6} 
 &
  Temperature &
  temp\_2m &
  Temperature at 2 m height   above surface &
   &
  (K) \\ \cline{2-6} 
 &
  Mean Sea Level Pressure &
  mslp &
  Atmospheric pressure at mean   sea level &
   &
  (Pa) \\ \hline
\multirow{5}{*}{Pressure Variables} & Geopotential Height & HGT\_prl\_\{plevel\} & Proportional to the height of   a pressure level & 1000, 850, 500, 50 & (m) \\ \cline{2-6} 
 &
  Relative Humidity &
  RH\_prl\_\{plevel) &
  Humidity relative to   saturation &
  850, 500 &
  (\%) \\ \cline{2-6} 
 &
  Temperature &
  TMP\_prl\_\{plevel\} &
  Temperature &
  850, 500 &
  (K) \\ \cline{2-6} 
 &
  U-Component of Wind &
  UGRD\_prl\_\{plevel\} &
  Wind in x/longitude-direction &
  850, 500 &
  (ms-1) \\ \cline{2-6} 
 &
  V-Component of Wind &
  VGRD\_prl\_\{plevel\} &
  Wind in y/latitude direction &
  850, 500 &
  (ms-1) \\ \hline
\end{tabular}%
}
\end{table}
\end{landscape}

\FloatBarrier

\begin{table}[p]
\centering
\small
\caption{Model details with different values of C}
\label{tab:Table2}
\vspace{10pt}
\resizebox{\columnwidth}{!}{%
\begin{tabular}{|l|l|l|l|l|l|}
\hline
C   & \begin{tabular}[c]{@{}l@{}}Total/Trainable\\    \\ Parameters\end{tabular} & Minimum validation loss & \begin{tabular}[c]{@{}l@{}}Forward/Backward\\    \\ pass size\\    \\ (MB)\end{tabular} & \begin{tabular}[c]{@{}l@{}}Parameters size\\    \\ (MB)\end{tabular} & \begin{tabular}[c]{@{}l@{}}Estimated total size\\    \\ (MB)\end{tabular} \\ \hline
192 & 23,849,296                                                                 & 0.098                   & 1627.12                                                                                 & 90.98                                                                & 1723.17                                                                   \\ \hline
96  & 6,017,200                                                                  & 0.072                   & 824.02                                                                                  & 22.95                                                                & 852.03                                                                    \\ \hline
48  & 1,531,744                                                                  & 0.080                   & 422.46                                                                                  & 5.84                                                                 & 433.37                                                                    \\ \hline
24  & 396,664                                                                    & 0.108                   & 221.68                                                                                  & 1.51                                                                 & 228.26                                                                    \\ \hline
12  & 106,036                                                                    & 0.162                   & 121.29                                                                                  & 0.40                                                                 & 126.76                                                                    \\ \hline
\end{tabular}%
}
\end{table}

\FloatBarrier

\begin{table}[p]
\centering
\small
\caption{Error matrices of all the selected variables at 6-hour lead time}
\label{tab:Table3}
\vspace{10pt}

\resizebox{\columnwidth}{!}{%
\begin{tabular}{|l|l|l|l|l|}

\hline
Variable       & RMSE   & ACC   & MAPE  & FSS   \\ \hline
uwind\_10m     & 0.800  & 0.972 & 3.040 & 0.868 \\ \hline
vwind\_10m     & 0.785  & 0.968 & 2.230 & 0.860 \\ \hline
temp\_2m       & 1.081  & 0.984 & 0.001 & 0.989 \\ \hline
mslp           & 61.012 & 0.995 & 0.000 & 0.967 \\ \hline
HGT\_prl\_1000 & 6.966  & 0.992 & 0.568 & 0.969 \\ \hline
HGT\_prl\_850  & 5.752  & 0.987 & 0.001 & 0.969 \\ \hline
HGT\_prl\_500  & 4.277  & 0.996 & 0.000 & 0.988 \\ \hline
HGT\_prl\_50   & 7.690  & 0.999 & 0.000 & 0.968 \\ \hline
RH\_prl\_850   & 6.196  & 0.945 & 0.066 & 0.936 \\ \hline
RH\_prl\_500   & 7.045  & 0.994 & 0.142 & 0.922 \\ \hline
TMP\_prl\_850  & 0.632  & 0.993 & 0.000 & 0.951 \\ \hline
TMP\_prl\_500  & 0.454  & 0.993 & 0.000 & 0.983 \\ \hline
UGRD\_prl\_850 & 1.025  & 0.972 & 3.832 & 0.904 \\ \hline
UGRD\_prl\_500 & 1.403  & 0.979 & 1.432 & 0.936 \\ \hline
VGRD\_prl\_850 & 0.955  & 0.948 & 4.039 & 0.871 \\ \hline
VGRD\_prl\_500 & 1.305  & 0.947 & 4.081 & 0.889 \\ \hline
\end{tabular}%
}
\end{table}


\begin{table}[p]
\centering
\caption{RMSE for different prediction approaches at 3 days/5 days lead time}
\label{tab:Table4}
\resizebox{\columnwidth}{!}{%
\begin{tabular}{|c|cc|cc|cc|}
\hline
\multirow{3}{*}{Variables} & \multicolumn{2}{c|}{Static}              & \multicolumn{2}{c|}{Autoregressive}      & \multicolumn{2}{c|}{Hierarchical}        \\ \cline{2-7} 
                           & \multicolumn{2}{c|}{RMSE}                & \multicolumn{2}{c|}{RMSE}                & \multicolumn{2}{c|}{RMSE}                \\ \cline{2-7} 
                           & \multicolumn{1}{c|}{(3days)}  & (5days)  & \multicolumn{1}{c|}{(3days)}  & (5days)  & \multicolumn{1}{c|}{(3days)}  & (5days)  \\ \hline
uwind\_10m                 & \multicolumn{1}{c|}{2.136359} & 2.210328 & \multicolumn{1}{c|}{1.68662}  & 1.997822 & \multicolumn{1}{c|}{1.606112} & 1.772361 \\ \hline
vwind\_10m                 & \multicolumn{1}{c|}{1.99046}  & 2.093402 & \multicolumn{1}{c|}{1.57443}  & 1.800885 & \multicolumn{1}{c|}{1.513184} & 1.633784 \\ \hline
temp\_2m                   & \multicolumn{1}{c|}{4.608834} & 4.675372 & \multicolumn{1}{c|}{2.303263} & 2.961949 & \multicolumn{1}{c|}{1.864674} & 2.228307 \\ \hline
mslp                       & \multicolumn{1}{c|}{346.6905} & 367.7903 & \multicolumn{1}{c|}{270.2304} & 328.7398 & \multicolumn{1}{c|}{246.9153} & 284.6909 \\ \hline
HGT\_prl\_1000             & \multicolumn{1}{c|}{34.0989}  & 35.36599 & \multicolumn{1}{c|}{26.24758} & 32.37408 & \multicolumn{1}{c|}{24.60023} & 27.77573 \\ \hline
HGT\_prl\_850              & \multicolumn{1}{c|}{26.88293} & 28.18798 & \multicolumn{1}{c|}{21.27528} & 25.33505 & \multicolumn{1}{c|}{19.9131}  & 22.90483 \\ \hline
HGT\_prl\_500              & \multicolumn{1}{c|}{25.22657} & 27.95444 & \multicolumn{1}{c|}{23.18886} & 25.88552 & \multicolumn{1}{c|}{21.75583} & 24.96543 \\ \hline
HGT\_prl\_50               & \multicolumn{1}{c|}{25.22396} & 29.94468 & \multicolumn{1}{c|}{25.27295} & 31.58878 & \multicolumn{1}{c|}{23.54968} & 26.97    \\ \hline
RH\_prl\_850               & \multicolumn{1}{c|}{16.73203} & 17.23171 & \multicolumn{1}{c|}{13.17513} & 15.46444 & \multicolumn{1}{c|}{12.24737} & 13.26609 \\ \hline
RH\_prl\_500               & \multicolumn{1}{c|}{19.48714} & 20.55311 & \multicolumn{1}{c|}{17.47008} & 18.8624  & \multicolumn{1}{c|}{15.91983} & 17.4563  \\ \hline
TMP\_prl\_850              & \multicolumn{1}{c|}{2.339094} & 2.507926 & \multicolumn{1}{c|}{1.906385} & 2.389406 & \multicolumn{1}{c|}{1.701259} & 2.092637 \\ \hline
TMP\_prl\_500              & \multicolumn{1}{c|}{1.815862} & 2.027463 & \multicolumn{1}{c|}{1.543275} & 1.821651 & \multicolumn{1}{c|}{1.473678} & 1.666765 \\ \hline
UGRD\_prl\_850             & \multicolumn{1}{c|}{3.123937} & 3.473405 & \multicolumn{1}{c|}{2.599124} & 3.212299 & \multicolumn{1}{c|}{2.462106} & 2.95791  \\ \hline
UGRD\_prl\_500             & \multicolumn{1}{c|}{4.565658} & 4.998262 & \multicolumn{1}{c|}{4.307457} & 5.050283 & \multicolumn{1}{c|}{3.779753} & 4.29788  \\ \hline
VGRD\_prl\_850 & \multicolumn{1}{c|}{2.796137} & 2.893178 & \multicolumn{1}{c|}{2.215406} & 2.496523 & \multicolumn{1}{c|}{2.163631} & 2.337054 \\ \hline
VGRD\_prl\_500             & \multicolumn{1}{c|}{4.466237} & 4.630708 & \multicolumn{1}{c|}{3.79047}  & 4.286071 & \multicolumn{1}{c|}{3.637562} & 4.016475 \\ \hline
\end{tabular}%
}
\end{table}
\clearpage  

\renewcommand{\floatpagefraction}{0.0}
\newpage

\begin{figure}[p]
\includegraphics[width=\textwidth]{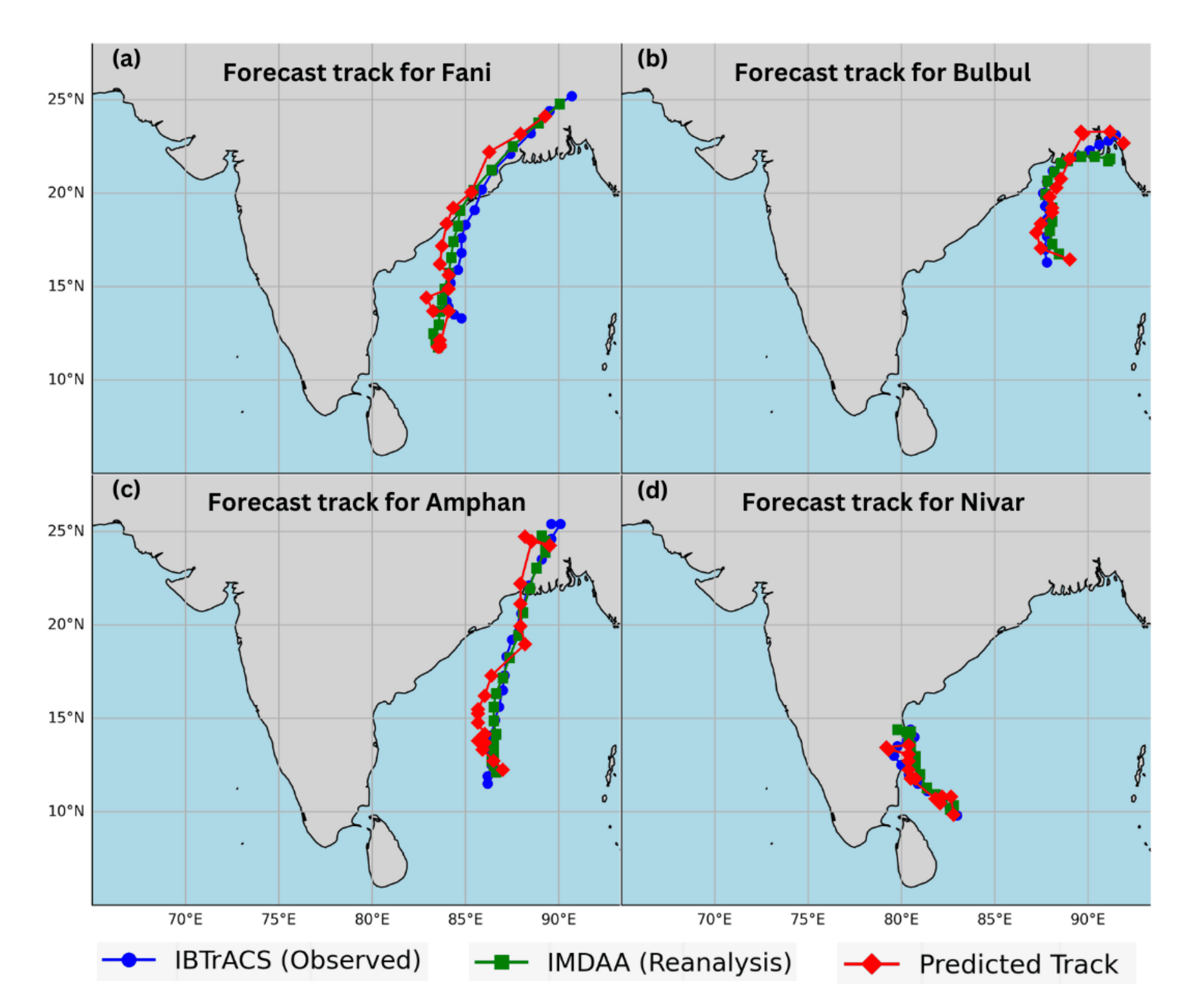}
\caption{Cyclone track comparison between predicted, observed, and reanalysis data for (a) Fani, (b) Bulbul, (c) Amphan, and (d) Nivar..
}
\label{fig:Figure6}
\end{figure}

\clearpage
\renewcommand{\figurename}{S}
\setcounter{figure}{0}
\renewcommand{\floatpagefraction}{0.0}
\newpage
\begin{figure}[p]
\includegraphics[width=\textwidth]{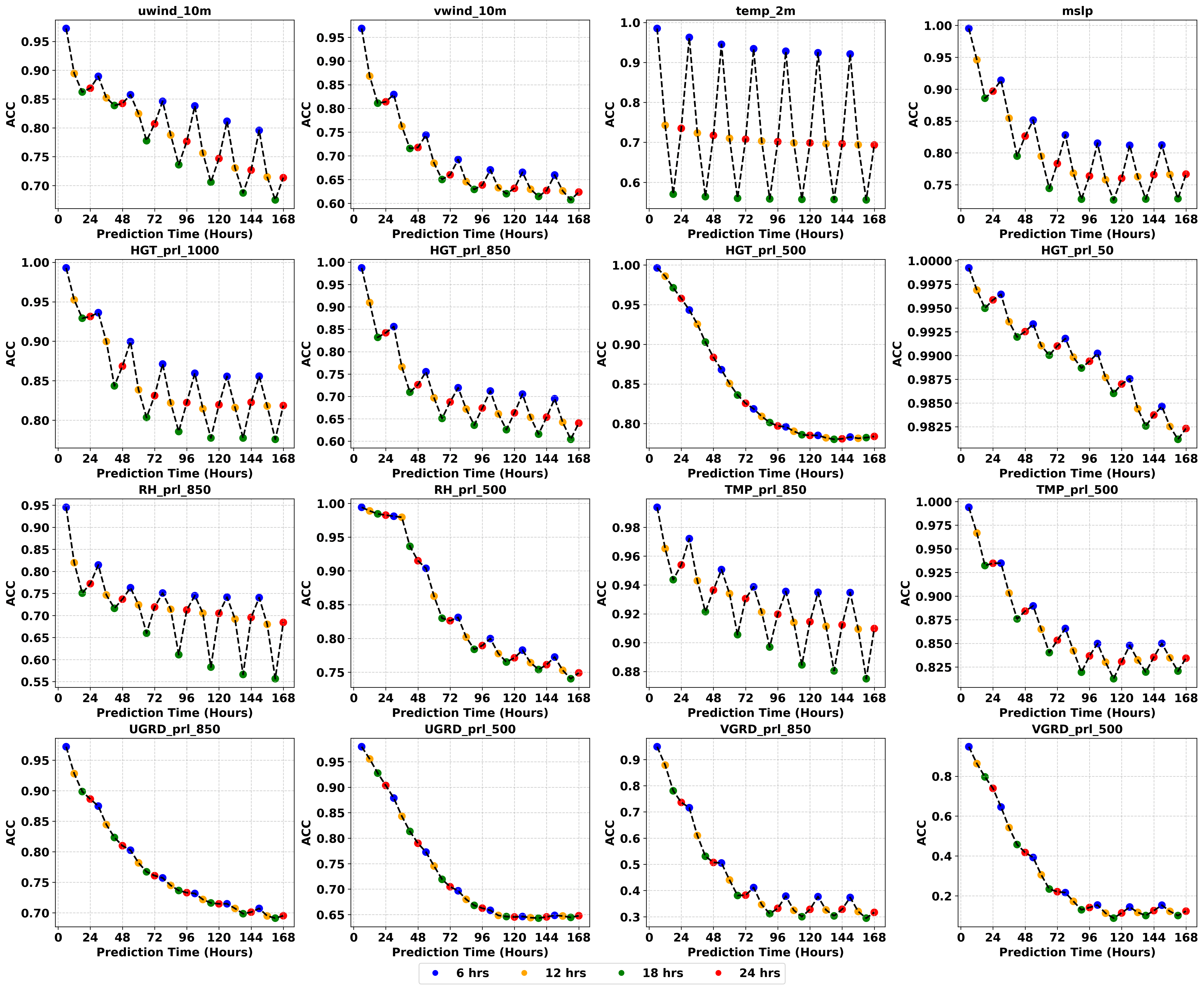}
\caption{ACC in predicting each variable for the next seven days with a static forecasting approach. The colors blue (06 hours), orange (12 hours), green (18 hours), and red (24 hours) represent the initial prediction and subsequent forecast at 24-hour intervals.
}
\label{fig:S1}
\end{figure}

\begin{figure}[p]
\includegraphics[width=\textwidth]{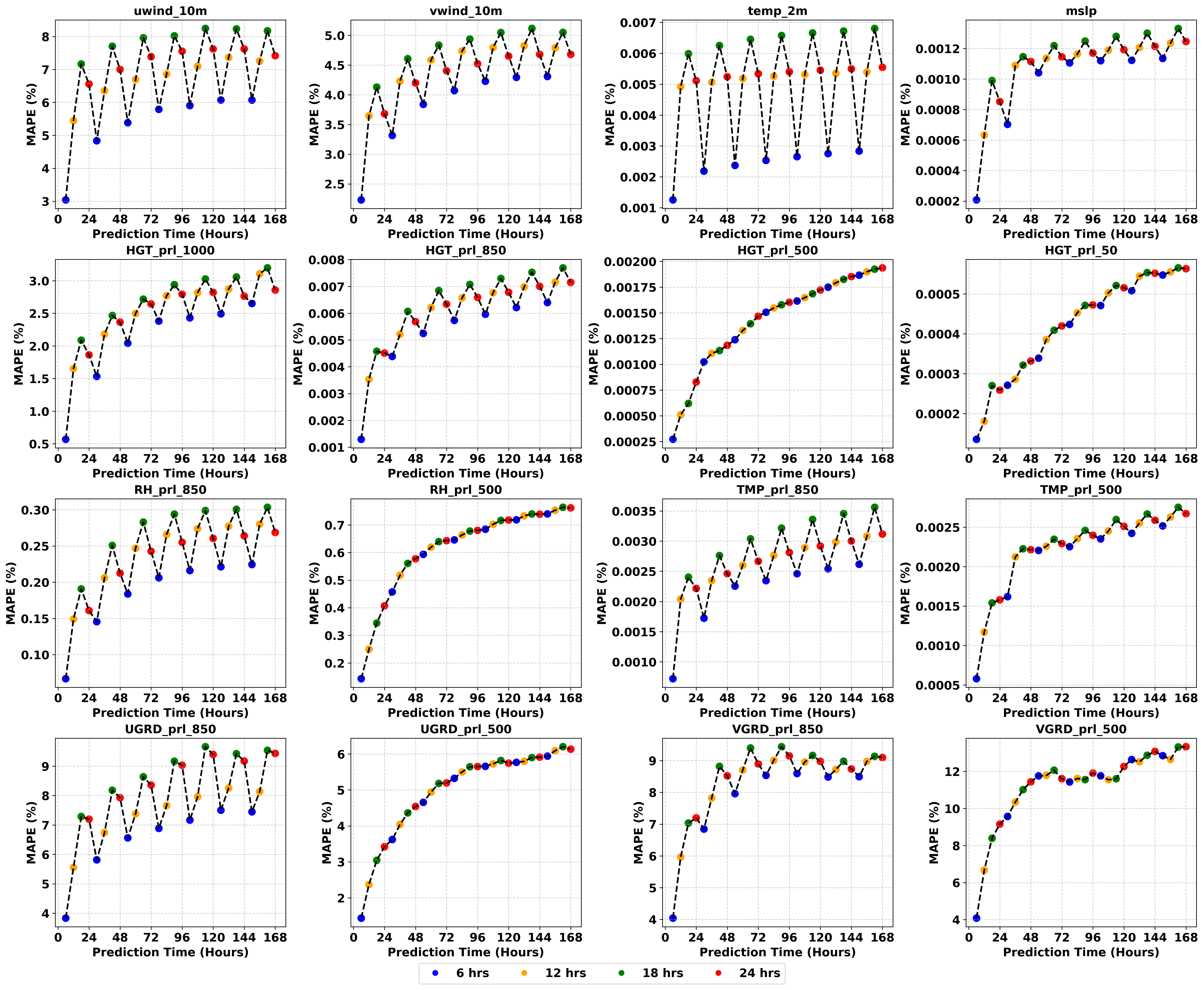}
\caption{MAPE in predicting each variable for the next seven days with a static forecasting approach. The colors blue (06 hours), orange (12 hours), green (18 hours), and red (24 hours) represent the initial prediction and subsequent forecast at 24-hour intervals.
}
\label{fig:S2}
\end{figure}

\begin{figure}[p]
\includegraphics[width=\textwidth]{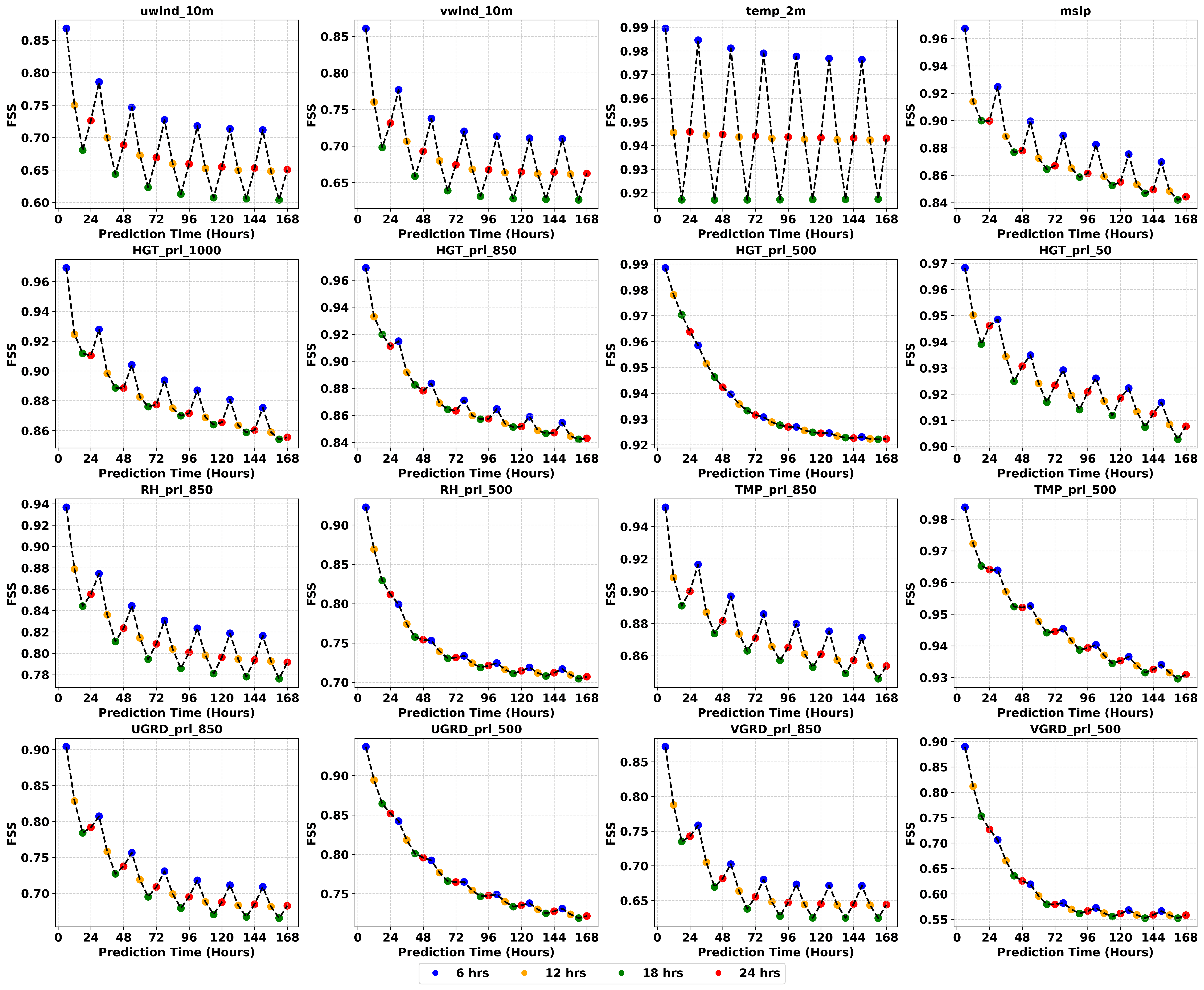}
\caption{FSS in predicting each variable for the next seven days with a static forecasting approach. The colors blue (06 hours), orange (12 hours), green (18 hours), and red (24 hours) represent the initial prediction and subsequent forecast at 24-hour intervals.
}
\label{fig:S3}
\end{figure}

\begin{figure}[p]
\includegraphics[width=\textwidth]{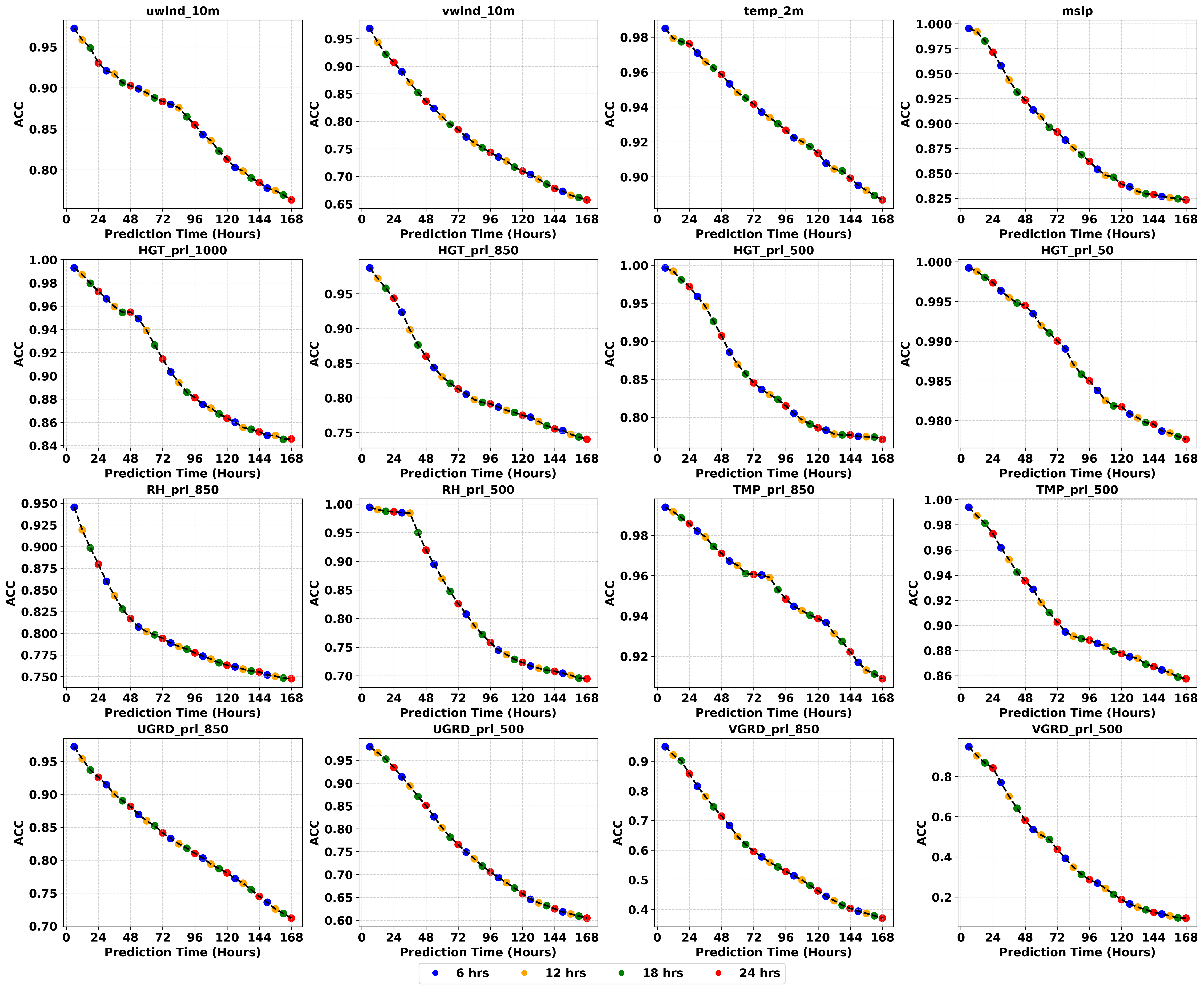}
\caption{ACC in predicting each variable for the next seven days with an autoregressive forecasting approach. The colors blue (06 hours), orange (12 hours), green (18 hours), and red (24 hours) represent the initial prediction and subsequent forecast at 24-hour intervals.
}
\label{fig:S4}
\end{figure}

\begin{figure}[p]
\includegraphics[width=\textwidth]{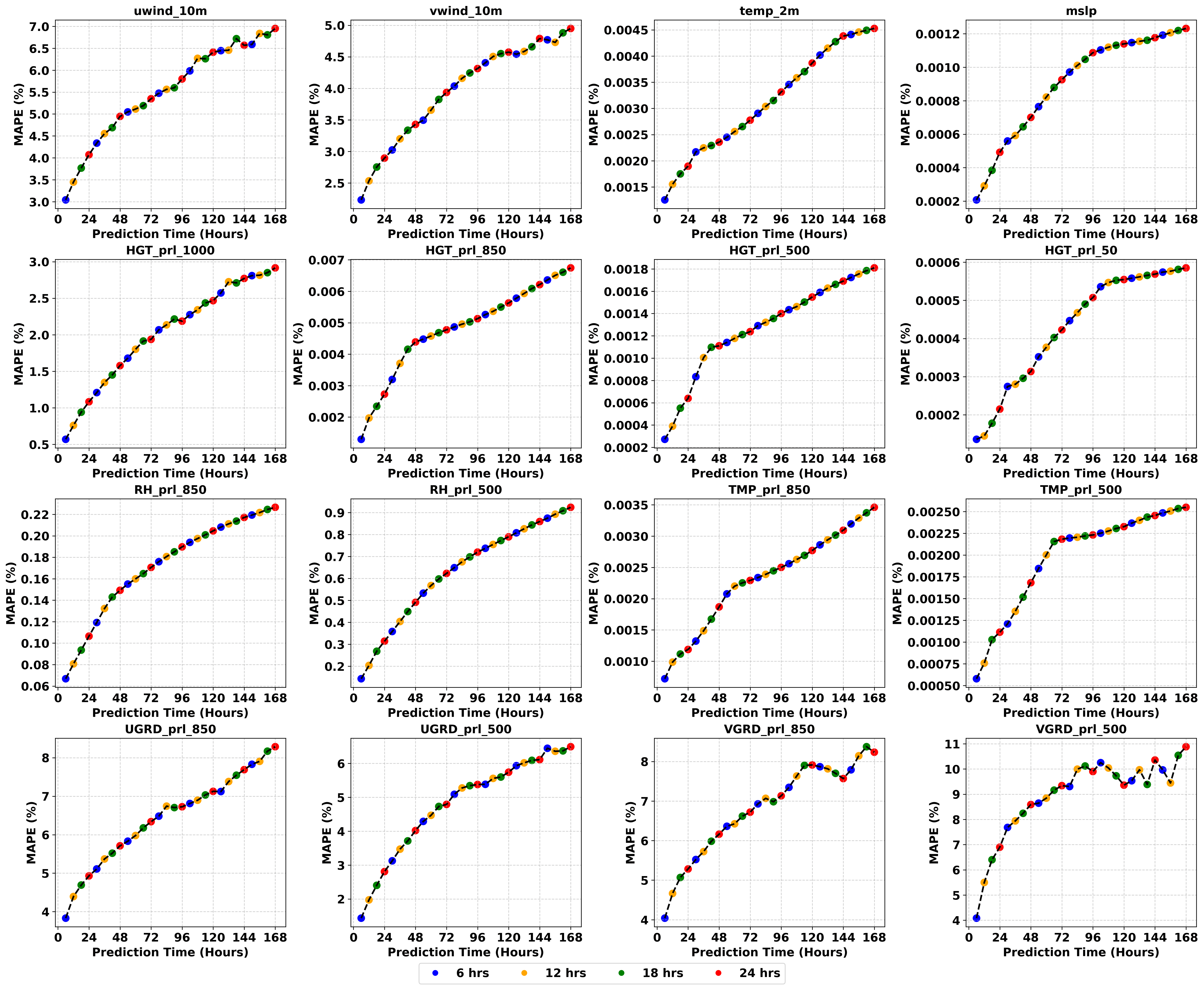}
\caption{MAPE in predicting each variable for the next seven days with an autoregressive forecasting approach. The colors blue (06 hours), orange (12 hours), green (18 hours), and red (24 hours) represent the initial prediction and subsequent forecast at 24-hour intervals.
}
\label{fig:S5}
\end{figure}

\begin{figure}[p]
\includegraphics[width=\textwidth]{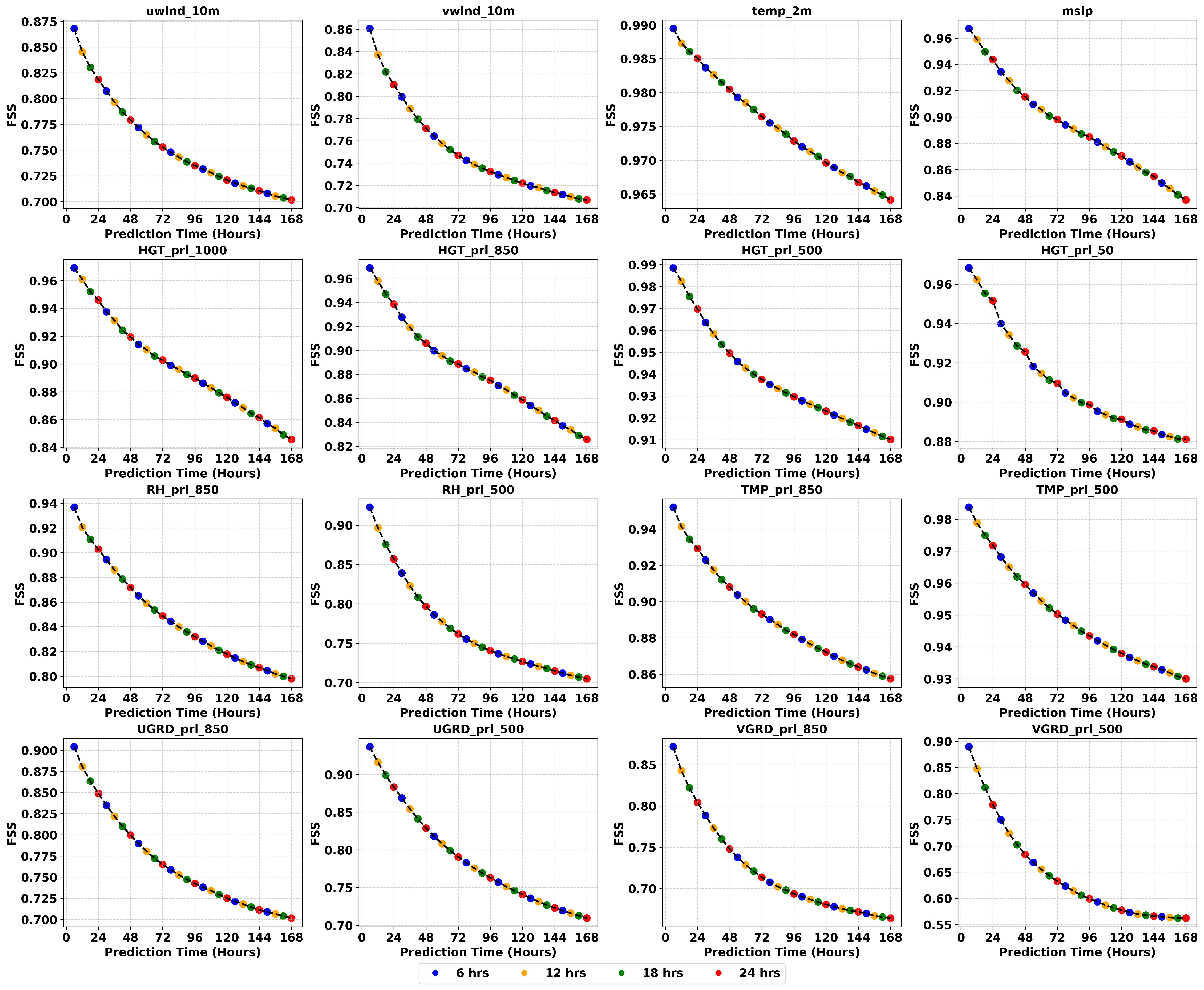}
\caption{FSS in predicting each variable for the next seven days with an autoregressive forecasting approach. The colors blue (06 hours), orange (12 hours), green (18 hours), and red (24 hours) represent the initial prediction and subsequent forecast at 24-hour intervals.
}
\label{fig:S6}
\end{figure}

\begin{figure}[p]
\includegraphics[width=\textwidth]{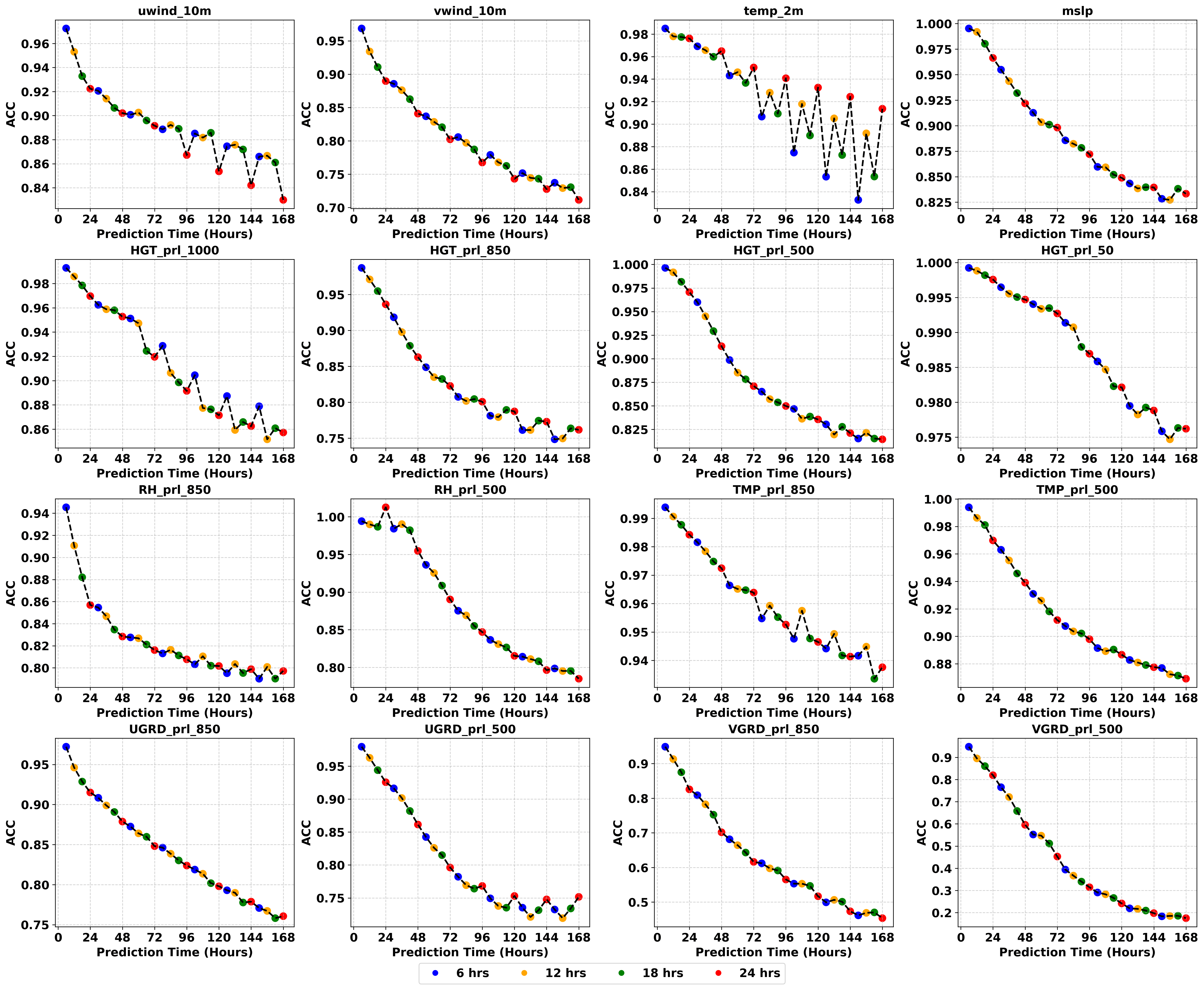}
\caption{ACC in predicting each variable for the next seven days with a hierarchical forecasting approach. The colors blue (06 hours), orange (12 hours), green (18 hours), and red (24 hours) represent the initial prediction and subsequent forecast at 24-hour intervals.
}
\label{fig:S7}
\end{figure}

\begin{figure}[p]
\includegraphics[width=\textwidth]{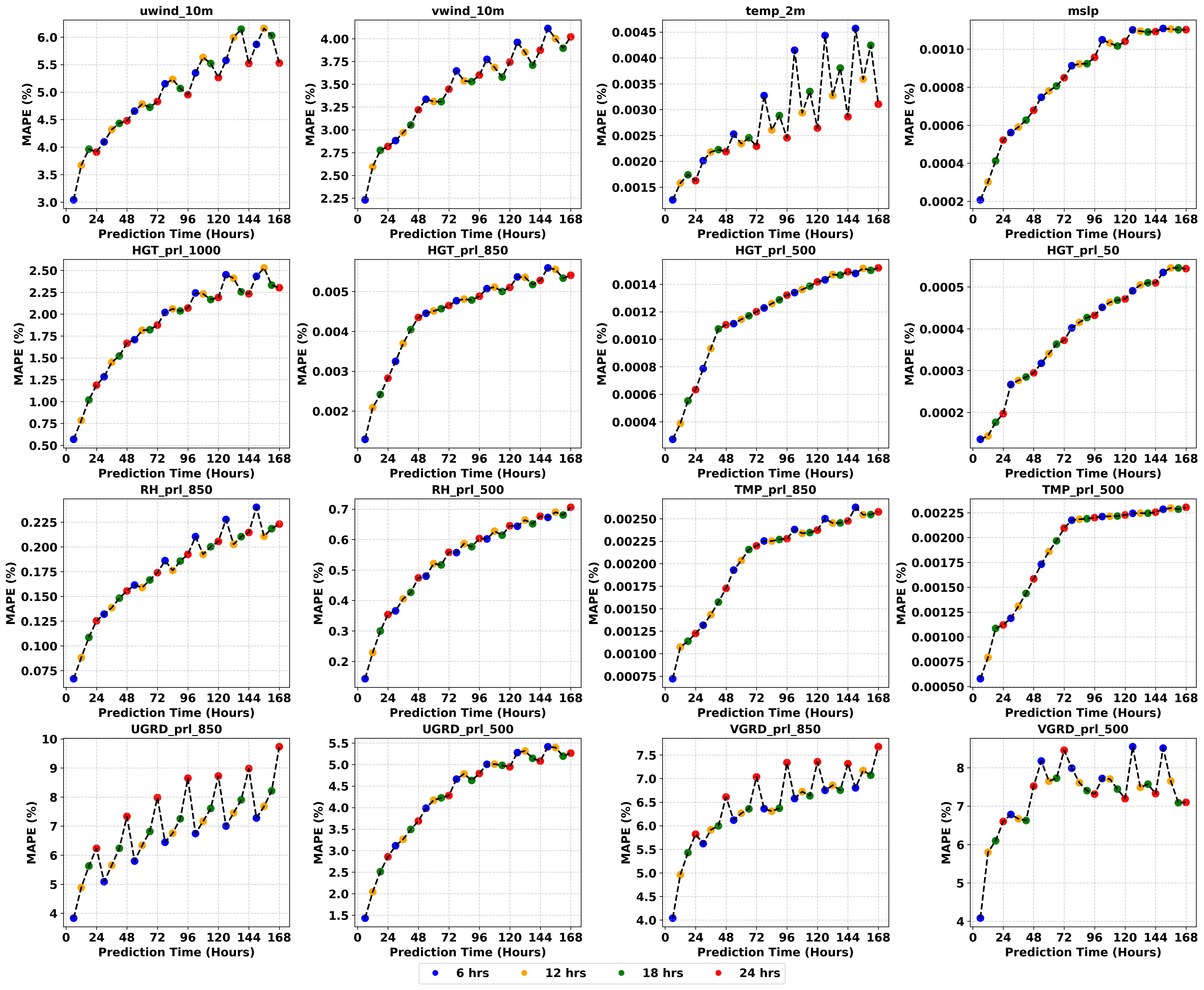}
\caption{MAPE in predicting each variable for the next seven days with a hierarchical forecasting approach. The colors blue (06 hours), orange (12 hours), green (18 hours), and red (24 hours) represent the initial prediction and subsequent forecast at 24-hour intervals.
}
\label{fig:S8}
\end{figure}

\begin{figure}[p]
\includegraphics[width=\textwidth]{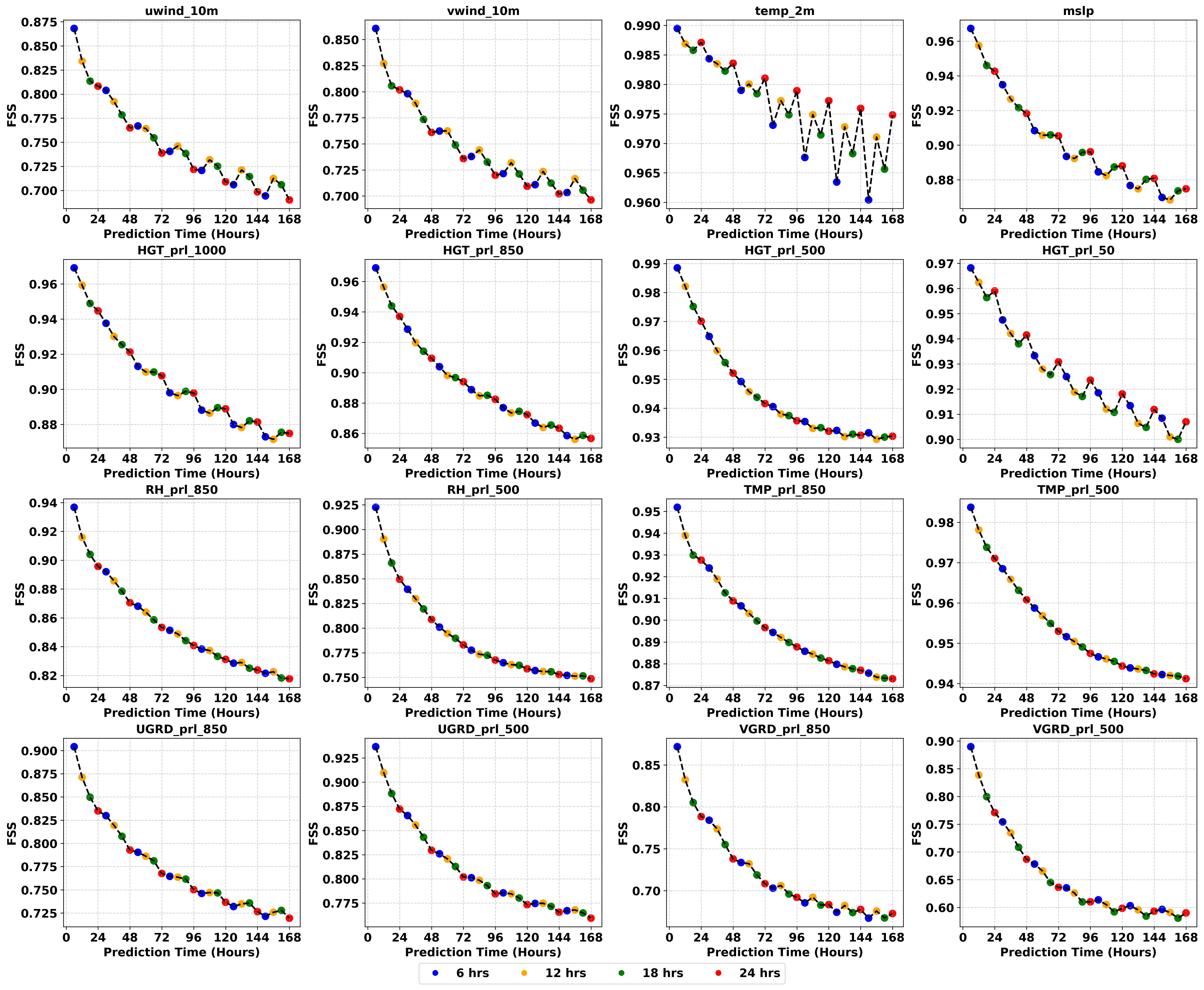}
\caption{FSS in predicting each variable for the next seven days with a hierarchical forecasting approach. The colors blue (06 hours), orange (12 hours), green (18 hours), and red (24 hours) represent the initial prediction and subsequent forecast at 24-hour intervals.
}
\label{fig:S9}
\end{figure}

\end{document}